\documentstyle[psfig,aps,12pt]{revtex}
\begin{document}
\title{Bose-Einstein condensation of indirect excitons in coupled quantum wells}
\author{G. M. Kavoulakis}
\address{Mathematical Physics, Lund Institute of Technology, P.O. Box 118,
        S-22100 Lund, Sweden \vspace{0.5in}\\}

\maketitle

\begin{abstract}

We study the ground-state properties of a quasi-two-dimensional
Bose-Einstein condensate of indirect excitons, which are confined in
an anisotropic harmonic potential. Incorporating the interactions, 
we calculate the order parameter variationally. The difficulties in
the detection of a Bose-Einstein condensate are also discussed,
along with possible ways which would overcome them.

\end{abstract}

\section{Introduction}
Excitons are bound states of electrons and holes which form in excited 
semiconductors \cite{Knox}, very much like positronium atoms. 
Since excitons consist of two fermions, at least as long as the mean
spacing between them is much larger than their Bohr radius, they are 
expected to behave as bosons \cite{Keldysh}. While the phase transition to a
Bose-Einstein condensed phase has been realized experimentally in vapours
of alkali-metal atoms \cite{PS}, the formation of an exciton Bose-Einstein
condensate (BEC) has turned out to be more tricky. A lot of
experimental effort has been made on the formation of a BEC of
excitons in Cu$_2$O \cite{AndreJim,Andre2}, and in quantum wells \cite{QW}.
Excitons resemble atoms, with one major difference being their mass,
since they are much lighter particles and as a result they exhibit
quantum-mechanical effects more easily. In that respect, excitons are more 
advantageous than atoms for forming a BEC. On the other hand, their 
basic disadvantage is their finite lifetime due to various decay
mechanisms \cite{Keith,KM}, which destroy them.

For a gas of bosons of mass $m$ and density $n$, the phase transition to a
Bose-Einstein condensed phase occurs when their thermal de-Broglie
wavelength becomes comparable to their interparticle distance, 
which implies that $k_B T_c \sim \hbar^2 n^{2/3}/m$, where $T_c$ is 
the critical temperature. For
$T_c$ being on the order of liquid-helium temperature, i.e., 1 K,
and for $m$ being on the order of the electron mass, $n$ turns out to 
be on the order of $10^{17}$ cm$^{-3}$. For such densities, the 
boson-boson spacing is $\sim 500$ \AA, while typically the exciton
Bohr radius is smaller than this length scale, and therefore the
excitons are expected to behave as point-like (bosonic) particles. 

Recently in an interesting paper Butov {\it et al.} \cite{Butov}
(see also \cite{Butov2}) reported the formation of a degenerate gas of 
excitons in a quasi-two-dimensional geometry in coupled quantum wells. More 
precisely, in the experiment of Ref.\,\cite{Butov} indirect excitons formed 
between two parallel layers, after an electric field was applied perpendicular 
to the two planes. Then, spatially resolving the photoluminescence that comes
from the layers, isolated bright localized spots were observed.
These spots were attributed to the recombining excitons which have
concentrated in regions of local minima (lakes) that exist due to
irregularities in the heterostructures. The presence of these local minima
is important, since the excitons concentrate in them. Furthermore, the
exciton lifetime is long due to the small overlap between the 
electron and the hole wavefunctions, and also the cooling due to the scattering 
of excitons with acoustic phonons is enhanced because of the relaxation of the 
momentum conservation along the direction of confinement. All the above effects
favor the formation of a cold and dense gas of excitons, and indeed signs of 
degenerate behaviour in these lakes were observed. 

Motivated by this experiment, we examine in the present paper the ground-state 
properties of a Bose-Einstein condensate of excitons in such a system 
\cite{PRS}. Our goal is to demonstrate some important physics which,
in addition to other things, would allow the detection of a Bose-Einstein 
condensate and not to give a detailed quantitative description. On the other 
hand, our results should be reliable at least as order-of-magnitude estimates. 
In that spirit, and since the real system is quite complicated, we make some 
simplifying assumptions. For example, we neglect the finite lifetime of 
excitons, since typically this is larger by one or two orders of magnitude
\cite{sup} than the thermalization timescale \cite{ILH}. In addition, the rate 
of non-radiative, Auger-like processes is expected to be suppressed because of 
the spatial separation between the electron and the hole of each individual 
exciton \cite{KBa}.

In what follows we examine in Sec.\,II the interactions between the
excitons and in Sec.\,III how the statistics due to the Pauli principle between
the electrons and/or the holes of different excitons affect them. We then make
some estimates in Sec.\,IV, where we model the in-plane confinement with a 
harmonic potential. Using a variational ansatz for the order parameter we 
calculate in Sec.\,V the energy and the width of the cloud in the case of an 
asymmetric confining potential. Finally we discuss in Sec.\,VI how one could 
get evidence for the existence of a BEC, while Sec.\,VII summarizes our
results. As shown, the width of the condensate 
is expected to be of the same order as that of the thermal cloud and for finite 
temperatures, that would obscure the detection of the condensate. In that 
respect, an important point of our study is that (similarly to the atomic 
condensates \cite{J}) as long as the trapping potential is {\it anisotropic}, 
the kinetic-energy distribution of an expanding BEC (upon release of the 
excitons from the trap) is also {\it anisotropic}, while the kinetic-energy
distribution of a thermal gas is always {\it isotropic}. Therefore, 
any experiment which is sensitive to the anisotropy of the kinetic-energy 
distribution of the excitons could be used as a ``smoking gun" for the 
detection of a BEC.

\section{Interaction between the excitons}
As mentioned above, the interactions play a very important role in
this system. Numerous studies have examined this problem - mostly 
the effect of the reduced dimensionality and of the band 
structure of specific materials. It is not the purpose
of the present paper to investigate such questions in detail,
but rather to follow a method that captures the essential physics
we want to demonstrate.

It is important to remember that the electrons and the holes in this
system are spatially separated and as a result the corresponding electron-hole
pairs (that form in the limit of dilute densities that we consider) have a
dipole moment. As shown in Ref.\,\cite{Yang} in such a system the effective
exciton-exciton interaction is dominated by the one that reduces to the 
usual dipole-dipole term for large exciton separations and this is the
approach that we follow here. Viewed in another way, this approach
is phenomenological. While it is certainly reliable
for large separations between the excitons, it is less reliable for short
distances, where one needs to worry about the exchange terms between the
electrons and the holes of different excitons. However, the dipole moment
of the excitons dominates the interaction and our approach is certainly
valid for low exciton densities; a similar approach was followed in
Ref.\,\cite{KU}. References \cite{at} have studied a problem close to
the present one in the context of atomic condensates which are subjected to 
an external electric field.

More precisely, let us consider the two-exciton problem, i.e., two electrons
confined on a plane and two holes confined on a parallel plane, with the planes
being separated by a distance $D$. As in the experiment of Ref.\,\cite{Butov} 
we assume that $D$ is on the order of the Bohr radius $a_B$ of a single exciton
(this is a reasonable thing to do, since if $D \sim a_B$
the electron-hole bound state is essentially a $p$-like hydrogenic state).
Fixing the position of, say, the two electrons at an infinite distance, in 
the lowest state each hole binds with one electron and two excitons form, with 
the total energy of the system being twice the binding energy of a single 
exciton. If the two excitons are brought to a finite distance from each other, 
the interaction between them is given by
\begin{equation}
 V({\bf r} - {\bf r}') = 
    \frac {e^2} {\epsilon} \left( \frac 2 {|{\bf r} - {\bf r}'|}
       - \frac 1 {|{\bf r} - {\bf r}' + {\bf D}|}
          - \frac 1 {|{\bf r} - {\bf r}' - {\bf D}|} \right),
\label{ei}
\end{equation}
where $\epsilon$ is the dielectric constant, and ${\bf r} - {\bf r}'$ is 
the vector connecting the electrons or the holes of the two excitons. 
In our study we assume for simplicity that the exciton dipole moment
is parallel to the electric field (i.e., perpendicular to the $x$-$y$
plane) and we neglect the fluctuations around this orientation.
For small distances between the two excitons, $|{\bf r} - {\bf r}'| \ll D$, 
Eq.\,(\ref{ei}) reduces to the usual Coulom interaction between the two
electrons and the two holes, $V({\bf r} - {\bf r}') = 2 e^2/(\epsilon |{\bf r}
- {\bf r}'|)$. For large distances between the two excitons, $|{\bf r} - 
{\bf r}'| \gg D$, Eq.\,(\ref{ei}) reduces to the usual dipole-dipole 
interaction, $V({\bf r} - {\bf r}') = 
 e^2 D^2 (1 - 3 \cos^2 \theta)/{\epsilon |{\bf r} - {\bf r}'|^3},$
where $\theta$ is the angle between ${\bf D}$ and ${\bf r} - {\bf r}'$.
In our case, since the system is quasi-two-dimensional, $\theta \approx \pi/2$.
Therefore $V({\bf r} - {\bf r}')$ is purely repulsive,
which guarantees that there are no instabilities against the formation
of e.g., molecules. 

\section{Effect of the Pauli principle}
When the separation between the excitons becomes of order $a_B$,
one has to worry about the Pauli principle between the electrons and/or
the holes. If the spins of either the electrons, or the holes are parallel,
the relative wavefunction between the two excitons is highly suppressed for
values of $|{\bf r} - {\bf r}'|$ smaller than $a_B$ (on the other hand,
this effect is absent when both the electrons and the holes have antiparallel
spins). As a result, the excitons are not allowed to get to distances shorter
than $|{\bf r} - {\bf r}'| \sim a_B$. Therefore, for exciton-exciton
separations on the order of $a_B$, the true many-body wavefunction behaves
in a similar way (since for the low densities we consider, the effect of 
other excitons is negligible) \cite{PSt}; however our mean-field approach
ignores these correlations at such small scales. For this reason, 
to calculate the interaction energy when the Pauli
principle is active, we will use Eq.\,(\ref{ei}) assuming that the two 
excitons cannot get to distances closer than $a_B$.  
It is important to point out two things concerning these implications
that result from the Pauli principle: (i) as shown in detail below, 
the energy of the system is not very sensitive to the choice of the cutoff 
length (which we choose to be $a_B$ for simplicity), and (ii) the bosonic 
nature of the electron-hole pairs is not affected, as long as the mean 
exciton-exciton spacing is larger than $a_B$, which is indeed the case in 
the experiment of Ref.\,\cite{Butov}.

\section{Estimates}
Before we calculate the order parameter of the condensate, it is 
instructive to make some estimates. Let us assume that the excitons
are confined on the $x$-$y$ plane within a lake of radius $d$.
For low enough densities the dipole expansion can be used and in this case
the energy per particle of a BEC of excitons is (apart from 
numerical factors of order unity),
\begin{eqnarray}
   {\cal E}(d) &\sim& \frac {\hbar^2} {2 m d^2} + \frac 1 2 m \omega^2 d^2
         + \frac {N e^2 D^2} {\epsilon d^2 D},
\label{en}
\end{eqnarray}
where $N$ is the number of excitons. In the above equation 
we have assumed that the excitons are trapped in a harmonic
potential of frequency $\omega$ (corresponding to the confinement
in the lake). Equation (\ref{en}) can also be written as
\begin{eqnarray}
  {\cal E}(d) \sim \frac {\hbar^2} {2 m d^2} \left( 1 + 
        \frac {N D} {a_B} \right) + \frac 1 2 m \omega^2 d^2,
\label{enn}
\end{eqnarray}
where $a_B = \epsilon \hbar^2 / \mu_x e^2 $ is the exciton Bohr radius,
with $\mu_x=m_e m_h/m$ being the reduced mass of the effective electron mass
$m_e$ and the effective hole mass $m_h$, and $m=m_e+m_h$. For
$\mu_x \sim 0.1 m_0$, where $m_0$ is the electron mass and $\epsilon \sim 10$,
then $a_B \sim 100$ \AA, which is $\sim D$. Under typical conditions, $N \sim
10^6$, the first term on the right side of Eq.\,(\ref{enn}) is negligible. 
Under this assumption, the confinement due to the lake 
is balanced by the (repulsive) interactions between the 
dipoles, while the kinetic energy (due to the Heisenberg uncertainty 
principle) is negligible, very much like the Thomas-Fermi approximation in
atomic condensates \cite{BP},
\begin{eqnarray}
   {\cal E}(d) \sim \frac {\hbar^2} {2 m d^2} \frac {N D} {a_B}
       + \frac 1 2 m \omega^2 d^2,
\label{est}
\end{eqnarray}
and thus the radius of the condensate is 
\begin{eqnarray}
       d_0/a_{\rm osc} = c_1 (N D / a_B)^{1/4},
\label{width}
\end{eqnarray}
with $a_{\rm osc} = (\hbar/m \omega)^{1/2}$ being the oscillator
length and $c_1$ being a constant of order unity. Also,
\begin{eqnarray}
   {\cal E}(d_0) / \hbar \omega = c_2 (N D /a_B)^{1/2},
\label{e0}
\end{eqnarray}
where $c_2$ is a constant of order unity.
A useful formula is the one that compares the radius of the
condensate $d_0$ with the radius of the thermal cloud $R$. Since
$m \omega^2 R^2 \sim k_B T$, and assuming that $k_B T_c \sim \hbar \omega
N^{1/2}$ \cite{PS},
\begin{eqnarray}
  \frac {d_0} R \sim \left( \frac {\hbar \omega} {k_B T} \right)^{1/2}
    \left( \frac {N D} {a_B} \right)^{1/4}
    \sim \left( \frac {T_c} T \right)^{1/2} \left( \frac {D} {a_B}
    \right)^{1/4}.
\label{dvr}
\end{eqnarray}
Remarkably, the ratio $d_0/R$ does not depend on the number of excitons $N$.
Turning to the speed of sound $c$, $m c^2 \sim \langle V \rangle$, where 
$\langle V \rangle \sim N e^2 D/ \epsilon d_0^2$ is the interaction energy 
per particle, and thus $c \sim (\hbar/m d_0) (N D/a_B)^{1/2}$. Also, 
the coherence or healing length $\xi$, satisfies the equation $\hbar^2 / 
2 m \xi \sim \langle V \rangle$, and thus $\xi/d_0 \sim (a_B/N D)^{1/2}$.
Finally the critical frequency $\Omega_c$ for creating a vortex state
is $\sim (\hbar/m d_0^2) \ln (d_0/\xi)$.

Let us make some numerical estimates now. Since the radius $R$ of a thermal
cloud of excitons in the lakes is $\approx 10$ $\mu$m for $T \approx 1$ K
\cite{Butov}, the formula $m \omega^2 R^2 \sim k_B T$ implies that $\omega \sim 
10^9$ Hz for $m$ being on the order of $0.1 m_0$. Given this value for
$\omega$, the oscillator length $a_{\rm osc}$ is $\sim 1$ $\mu$m, and $\hbar 
\omega \sim 5 \times 10^{-4}$ meV. Therefore, since $D \sim a_B$, then $N D/a_B 
\sim N \sim 10^6$, and the size of the condensate $d_0$ is $\sim 30$ $\mu$m, 
the surface density $n_{2D}$ is $\sim 3 \times 10^{10}$ cm$^{-2}$ and the mean 
exciton-exciton spacing $\sim d_0/N^{1/2}$ is $\sim 300$ \AA. The fact that 
$n_{2D} \, a_B^2$ is $\sim 0.03$, i.e., much less than unity,
where $n_{2D}$ is the surface density, justifies the treatment of the excitons
as bosons and the use of Eq.\,(\ref{ei}) as the appropriate expression for the
interaction between the excitons. Turning to the energy per particle of the
condensate, this is $\sim 1$ meV. The transition temperature $T_c$ is
$\sim$ 5 K, while for $T=T_c$, Eq.\,(\ref{dvr}) implies that $d_0 \approx R$.
Also the speed of sound $c$ is $\sim 10^6$ cm/s, and the coherence length
$\xi$ is $\sim 100$ \AA, which gives the typical size of vortices \cite{Butov2}.
Since $\xi/d_0 \sim 10^{-3}$, the superfluid properties of this system should
resemble those of helium (and not those of nuclei). Finally the critical 
frequency $\Omega_c$ for creating a vortex state is $\sim 5 \times 10^5$ Hz.

\section{Variational calculation of the order parameter}
Let us now get more precise. If $\Psi({\bf r})$ is the order parameter of
a BEC of excitons, as long as the estimates made earlier are accurate, it
is enough to consider only the relevant terms in the Hamiltonian,
\begin{eqnarray}
	 \left( V_t({\bf r}) + \int V({\bf r} - {\bf r}')
	  |\Psi({\bf r}')|^2 \, d^2 r' \right) \Psi({\bf r}) =
             \mu \Psi({\bf r}),
\label{ops}
\end{eqnarray}
where $V_t({\bf r})$ is the trapping potential and $\mu$ is the chemical
potential. We will now use the variational approach to calculate
the order parameter and the energy. For simplicity we restrict
ourselves to two dimensions \cite{2d}, i.e., on the $x$-$y$ plane
and assume that the trapping potential has the form
\begin{eqnarray}
   V_t({\bf r}) = \frac 1 2 m \omega^2 (x^2 + \lambda^2 y^2),
\end{eqnarray}
with $\omega = \omega_x$, and $\lambda = \omega_y/\omega_x$, and also
\begin{eqnarray}
     \Psi({\bf r}) = \frac {N^{1/2}} {\pi^{1/2} (a_x a_y)^{1/2}}
            \exp(-x^2/2 a_x^2-y^2/2 a_y^2),
\label{trf2d}
\end{eqnarray}
where $a_x$ and $a_y$ are variational parameters. Then, the energy per particle
is,
\begin{eqnarray}
  {\cal E} = \frac 1 N \left( \int V_t({\bf r}) |\Psi({\bf r})|^2 \, d^2 r 
     + \frac 1 2 \int V({\bf r} - {\bf r}')
 |\Psi({\bf r})|^2 |\Psi({\bf r}')|^2
 \, d^2 r \, d^2 r' \right).
\label{a1}
\end{eqnarray}
Starting with the expectation value of $V_t$,
$\langle \Psi |V_t| \Psi \rangle/N = m \omega^2 (a_x^2 + \lambda^2
a_y^2)/4$. Turning to the interaction energy
\begin{eqnarray}
      \langle V \rangle =
     \frac {e^2} {2 N \epsilon} \int |\Psi({\bf r})|^2 |\Psi({\bf r}')|^2
            \left( \frac 2 {|{\bf r} - {\bf r}'|}
	  - \frac 1 {|{\bf r} - {\bf r}' + {\bf D}|}
	 - \frac 1 {|{\bf r} - {\bf r}' - {\bf D}|}
	 \right) \,  d^2 r \, d^2 r',
\label{ddpo}
\end{eqnarray}
where, as we argued earlier, $|{\bf r} - {\bf r}'| \ge a_B$.
To evaluate the above integral, we use the relative coordinates
${\bf r}_{12} = {\bf r} - {\bf r}'$, and ${\bf R}_{12}=({\bf r}+{\bf r}')/2$.
Equation (\ref{ddpo}) then involves an integration over ${\bf R}_{12}$ and 
${\bf r}_{12}$,
\begin{eqnarray}
      \langle V \rangle =
            \frac {N e^2} {\pi^2 \epsilon ({a}_x {a}_y)^2}
	    \int e^{-2 R_{12}^2 f(\phi)} \, d^2{R_{12}}  
    \int \frac 1 {r_{12}} \left( 1 - 
     \frac {r_{12}}{\sqrt{{r_{12}^2} + D^2}} \right)
        e^{- r_{12}^2 f(\phi')/2} \, d^2r_{12},
\label{divintf}
\end{eqnarray}
where $f(\phi) = (\cos \phi / a_x)^2 + (\sin \phi / a_y)^2$.
Since $D \ll a_x, a_y$, we take the exponential in the integral over 
${\bf r}_{12}$ to be equal to unity, and thus
\begin{eqnarray}
  \int_{a_B}^{\infty} \frac 1 {r_{12}} \left( 1 - \frac {r_{12}} 
  {\sqrt{{r_{12}^2} + D^2}} \right)
          e^{- r_{12}^2 f(\phi')/2} \, d^2r_{12}
    &\approx& \int_{a_B}^{\infty} \frac 1 {r_{12}} \left( 1 - \frac {r_{12}} 
	{\sqrt{{r_{12}^2} + D^2}} \right)
	\, d^2r_{12} 
\nonumber \\
	&=& 2 \pi (\sqrt{a_B^2 + D^2} - a_B).
\label{simple}
\end{eqnarray}
Combining the equations above, the energy per particle is
\begin{eqnarray}
   {\cal E} =  \frac 1 4 m \omega^2 (a_x^2 + \lambda^2 a_y^2) 
  + \frac {2 N e^2 (\sqrt{a_B^2 + D^2} - a_B)} {\pi \epsilon (a_x a_y)^2} 
    \int e^{-2 R_{12}^2 f(\phi)} \, d^2R_{12}
\label{envafin}
\end{eqnarray}
expressed in terms of the two variational parameters $a_x, a_y$, 
and $\lambda$. For $\lambda =1$, then $a_x=a_y=d$, $f=1/d^2$, the integral in 
Eq.\,(\ref{envafin}) is equal to $\pi d^2/2$, and Eq.\,(\ref{envafin}) gets 
the simple form
\begin{eqnarray}
   {\cal E} &=&  \frac 1 2 m \omega^2 d^2 + \frac {N e^2 (\sqrt{a_B^2 +
   D^2} - a_B)} {\epsilon d^2}.
\label{envafins}
\end{eqnarray}
As Eq.\,(\ref{envafins}) implies, the energy of the system scales as
$(\sqrt{a_B^2 + D^2} - a_B)^{1/2}$, which shows that indeed ${\cal E}$
is not very sensitive to the cutoff length for $a_B \sim D$. Furthermore,
if the cutoff length in Eq.\,(\ref{simple}) is set equal to zero, 
\begin{eqnarray}
   {\cal E} &=&  \frac 1 2 m \omega^2 d^2 + \frac {N e^2 D} {\epsilon d^2},
\label{envafinss}
\end{eqnarray}
in agreement with the estimates of Eq.\,(\ref{est}). 
Equation (\ref{envafinss}) implies that $c_1$ in
Eq.\,(\ref{width}) is equal to $(2 m/\mu_x)^{1/4}$, and 
$c_2$ in Eq.\,(\ref{e0}) is equal to $(2m/\mu_x)^{1/2}$, where $m=m_e+m_h$.

Returning to the more general problem of $\lambda \neq 1$, we have
minimized the energy of Eq.\,(\ref{envafin}), setting for simplicity
the cutoff length equal to zero. Measuring $a_x$
and $a_y$ in the second term of this expression in units of $a_{\rm osc}$,
the energy scale $2 N e^2 D/\pi \epsilon a_{\rm osc}^2$ can be written as
$\hbar \omega (2 N D/\pi a_B) (m/\mu_x)$. We choose $(2 N D/\pi a_B)
(m/\mu_x)$ to be equal to $0.5 \times 10^6$. We also consider the value 
$\lambda=0.8$, which corresponds to a trapping potential with equipotential 
lines which are elongated along the $y$ axis, with a ratio of the oscillator 
lengths $a_{{\rm osc},x}/a_{{\rm osc},y}$ equal to $\lambda^{1/2} \approx 
0.894$. Minimizing numerically the energy, we find that ${\cal E}/\hbar \omega
\approx 1.119 \times 10^3$ with $a_x/a_{\rm osc} \approx 33.3$ and ${a}_y/
a_{\rm osc} \approx 42.0$. It is interesting that the ratio $a_x/a_y \approx 
0.793$ is smaller than $\lambda^{1/2}$. Therefore, the interactions enhance the 
anisotropy of the cloud due to the trapping potential, (i.e., they make the 
interacting cloud more elongated along the $y$ axis), as compared
to a non-interacting gas, which would result from the uncertainty principle. 
A similar effect has been observed in atomic condensates with contact
interactions \cite{J,BP}.

\section{Discussion of the results and experimental relevance}
In the last part of this study we examine how the above results can be 
used in practice. Getting indisputable evidence and clear signatures
for the existence of a BEC of excitons is not at all a trivial problem
\cite{JK}, since even for small but finite temperatures the condensate 
co-exists with the thermal cloud. More precisely, as Eq.\,(\ref{dvr}) implies,
the width of the condensate is, even for $T=T_c$, comparable or even larger 
than that of the thermal component, and this ratio gets even larger as the 
temperature decreases. This is opposite to the atomic condensates, where 
typically the thermal cloud is much broader than the condensate.
This fact would certainly obscure the detection of a BEC of excitons in
the present system.

In the alkali-metal atoms, in the
first experiment where the formation of a BEC was reported \cite{J}, the atoms 
were confined in a disk-shaped trap. Following their release from the
trap, for very low -- essentially zero -- temperatures, the kinetic-energy 
distribution of the atoms was observed to be anisotropic, while for higher 
temperatures (but still lower than $T_c$) a more broad isotropic background 
was also observed, corresponding to the thermal component. Above $T_c$, only 
the isotropic component remained. Therefore, the anisotropic expansion of the
cloud is a {\it unique} characteristic of a BEC, and {\it does not} show up in
a thermal cloud.

Turning to the present problem, ideally one could think about releasing 
somehow the excitons from the lake and watching them as they expand. Actually,
such a possibility has already been studied theoretically \cite{GP}.
In such an experiment, any sign of anisotropic expansion of the cloud on the
$x$-$y$ plane below some temperature (the transition temperature $T_c$) would
provide strong evidence in favour of the presence of a BEC of excitons.
On the other hand, from the estimates made earlier, one can see that the
typical velocity of expansion is expected to be $\sim ({\cal E}/m)^{1/2} \sim
10^6$ cm/s. Since excitons with velocities up to $\approx 1.4 \times
10^6$ are radiatively active \cite{ra}, the expanding excitons may or may not
be radiatively active. Additionally, since the velocity of expansion is close
to the sound velocity of the material ($\approx 3.7 \times 10^5$ cm/s), 
the interaction with phonons will slow them down. Apart from these possible
difficulties, more generally one should remember that any experiment that is 
sensitive to the directionality of the kinetic-energy distribution of the 
excitons could be used to provide evidence for the formation of a BEC.

\section{Summary}
To summarize, in this study we examined the ground-state properties of 
a quasi-two-dimensional Bose-Einstein condensate of indirect excitons, 
which are confined in an anisotropic harmonic potential. Incorporating the
interactions we made estimates for the size of the cloud, the energy 
per particle, the density, the speed of sound, the coherence length, and the
critical frequency for creating a vortex state. We also calculated
the order parameter variationally within the mean-field approximation and we 
examined the effect of the Pauli principle. Concerning the detection of 
a BEC, since the size of the condensate is expected to be of the same order as
that of the thermal cloud, we examined the possibility of releasing the
excitons from an anisotropic trap, which would result in an anisotropic
kinetic-energy distribution and it would provide evidence for the presence of 
a BEC. An interesting result is that the interacting gas exhibits this
anisotropy in a more pronounced way as compared to an ideal gas.

\section{Acknowledgments}
The author is grateful to G. Baym, A. D. Jackson, I. E. Perakis,
and S. M. Reimann for useful discussions. This work was supported 
by the Swedish Research Council (VR), and by the Swedish Foundation for
Strategic Research (SSF).

\end{document}